\documentclass[10pt,aps,prd]{revtex4-1}
\usepackage{amsfonts,amsmath,amssymb}
\usepackage{array,latexsym,mathrsfs}
\usepackage{url,hyperref}
\usepackage{orcidlink}
\usepackage{color}

\def\be{\begin{equation}}
\def\ee{\end{equation}}
\def\bs{\begin{split}}
\def\es{\end{split}}
\def\bea{\begin{eqnarray}}
\def\eea{\end{eqnarray}}
\def\ba{\begin{aligned}}
\def\ea{\end{aligned}}
\def\nn{\nonumber}

\def\pd{\partial}
\def\ta{\tilde{a}}
\def\tm{\tilde{m}}
\def\tn{\tilde{n}}
\def\ts{\tilde{s}}
\def\tr{\tilde{r}}
\def\tx{\tilde{x}}

\def\talpha{\tilde{\alpha}}
\def\tD{\widetilde{\Delta}}
\def\tP{\widetilde{P}}
\def\tX{\tilde{X}}
\def\tY{\tilde{Y}}
\def\tsigma{\tilde{\sigma}}
\def\tSigma{\widetilde{\Sigma}}
\def\ts{\tilde{s}}
\def\bs{\bar{s}}
\def\bt{\bar{t}}
\def\bp{\bar{\phi}}
\def\tp{\tilde{\phi}}
\def\Tr{\mathop{\rm Tr}\nolimits}
\def\diag{\mathop{\rm diag}\nolimits}
\def\bichi{\chi\!\!\!\!\chi}
\def\bg{\mathbf{g}}
\def\bm{\mathbf{m}}
\def\bn{\mathbf{n}}
\def\bl{\mathbf{l}}
\def\bU{\mathbf{U}}
\def\bV{\mathbf{V}}
\def\bW{\mathbf{W}}
\def\bA{\mathbf{A}}
\def\bGa{\mathbf{\Gamma}}
\def\bPsi{\mathbf{\Psi}}

\def\bx{\bar{x}}
\def\by{\bar{y}}

\begin{document}

\title{Is the type-D NUT C-metric really ``missing" from the most
general Pleba\'nski-Demia\'nski solution?}

\author{Shuang-Qing Wu\orcidlink{0000-0001-7936-7195}}
\email{sqwu@cwnu.edu.cn}

\author{Di Wu\orcidlink{0000-0002-2509-6729}}
\email{Contact author: wdcwnu@163.com}
\affiliation{School of Physics and Astronomy, China West Normal University,
Nanchong, Sichuan 637002, People's Republic of China}

\date{Received 27 August 2024; Accepted 21 October 2024}

\begin{abstract}
It remains a long-standing problem, unsettled for almost two decades in the general relativity community,
ever since Griffiths and Podolsk\'{y} demonstrated in their previous paper [J.B. Griffiths and J.
Podolsk\'{y}, \href{http://dx.doi.org/10.1088/0264-9381/22/17/008}{Classical Quantum Gravity
\textbf{22}, 3467 (2005)}] that the type-D NUT C-metric seems to be absent from the most general
family of the type-D Pleba\'nski-Demia\'nski (P-D) solution. However, Astorino, in his recent
article [M. Astorino, \href{https://doi.org/10.1103/PhysRevD.109.084038}{Phys. Rev. D \textbf{109},
084038 (2024)}] presented a different form of rotating and accelerating black holes and showed that
all known four-dimensional type-D accelerating black holes (without the NUT charge) can be recovered
via various different limits in a definitive fashion. In particular, he provided, for the first time,
the correct expressions for the type-D static accelerating black holes with a nonzero NUT charge,
which was previously impossible using the traditional parametrization of the familiar P-D solution.
Nevertheless, it still remains elusive how these two different forms of the four-dimensional
rotating and accelerating solutions are related. In this paper, we aim to fill this gap by finding
the obvious coordinate transformations and parameter identifications between the vacuum metrics
after two different parametrizations of the generated solution via the inverse scattering method
from the seed metric -- the Rindler vacuum background. We then resolve this ``missing" puzzle by
providing another M\"{o}bius transformation and linear combinations of the Killing coordinates,
which clearly cast the type-D NUT C-metric into the familiar form of the P-D solution. Additionally,
we propose an alternative new routine for the normalization of the obtained metric derived via the
inverse scattering method from the vacuum seed solution, which could be potentially useful for the
construction of higher-dimensional solutions using the trivial vacuum background as the seed metric.
\end{abstract}

\maketitle

\section{Introduction}

The most general type-D solution in general relativity has been well known for more than half a
century \cite{JMP10-1195,AP90-196,AP98-98}, and is often referred to as the Pleba\'nski-Demia\'nski
(P-D) metric. With the advent of the conveniently factorized form \cite{CQG22-109} of two quartic
structure functions, this solution has subsequently been extensively studied during the past twenty
years, see the book \cite{GP2009} for a comprehensive review. However, as demonstrated in Ref.
\cite{CQG22-3467}, the type-D NUT C-metric is \emph{apparently} accelerating and can be transformed
into the usual Taub-NUT solution. Since then, ``such a solution has to date neither been identified
nor proved not to exist." This raised a ``missing" puzzle for almost two decades, as the type-D
NUT C-metric appears to be absent from the most general family of the type-D P-D solution.

Recently, Astorino \cite{PRD109-084038} presented a different form of the rotating and accelerating
black holes that belong to the type-D family (although no obvious proof of this issue is provided
in that paper) and showed that all known four-dimensional type-D accelerating black holes without
the NUT charge can be obtained via various different limits in a definitive fashion. Moreover, he
presented, for the first time, the correct expressions for the type-D NUT C-metric, which are not
available in the traditional parametrization for the familiar P-D solution. Lately, this solution
has also been generalized in \cite{2404.06551} to include the electromagnetic charges and a nonzero
cosmological constant.

As far as the NUT C-metric is concerned, apart from the above type-D one, there is also another kind
of accelerating Taub-NUT solution \cite{PRD74-084031}, which belongs to a more general type I
\cite{PRD102-084024} that has attracted some recent interest \cite{GBoldi,PRD108-124025,PRD108-024059,
JHEP0823085,GRG56-111}. Although the initial form of this solution is complicated and has been obtained
via a hybrid solution-generating method from a specific seed, in fact it can be simply generated
\cite{GBoldi,PRD108-124025,PRD108-024059} via the Ehlers transformation from the C-metric. The Ehlers
transformation not only endows the black hole horizon with the NUT charge, but also allocates a
port of the NUT charge to the accelerating horizon. From the perspective of the double Kerr-Schild
formalism, the type-D NUT C-metric can be viewed as a perturbation on the pure Rindler background.
It seems that one could regard the NUT-charged Rindler metric as the background spacetime of the
type-I NUT C-metric. However, it is unclear whether the type-I C-metric admits a similar double
Kerr-Schild formulation. Based on the same reasoning, when the Harrison transformation is applied
to the seed C-metric, one gets a type-I electrically charged C-metric, whose background is the
electrified Rindler vacuum. However, a well-known type-D charged static accelerating solution
(Reissner-Nordstr\"{o}m C-metric, or RN C-metric) is already included in the most general P-D family.
Like the Ehlers transformation that allocates a port of the NUT charge to the Rindler background,
the Harrison transformation also assigns a port of the electric charge to the Rindler background.
Thus, in the type-I RN C-metric, the accelerating horizon acquires the electric charge; in other
words, its background is a charged Rindler metric. In this sense, the original Kinnersley-Walker's
interpretation \cite{PRD2-1359} of the uniformly accelerating charged mass as two ``oppositely"
accelerated black holes apart from each other seems not to be very accurate, as it cannot
distinguish the type-D accelerating black holes from the type-I ones. Rather, one could say
that the type-D accelerating solution is a black hole sitting on the pure Rindler background,
while the type-I accelerating metric is that on the more general Rindler background with dynamical
parameters (the NUT charge, electromagnetic charge) plus other kinetic parameters (such as the
rotation parameter and possibly the cosmological constant). It is noted that while the type-D
solution admits an extension to include a nonzero cosmological constant, the type-I metric seems
unable to permit such a generalization. By the way, we mention that there is also another kind of
NUT C-metric presented in Ref. \cite{PRD91-124005}, which is dubbed the [4-1]-soliton solution
\cite{PRD93-044021}. This solution still remains unexplored to date, but we can anticipate it to
be of type I in the Petrov classification.

What remains opaque in the Astorino's work is how these two different forms of the four-dimensional
rotating and accelerating solutions are related, and why the type-D NUT C-metric is ``missing" from
the P-D metric. The main purpose of this paper is to resolve this ``missing" puzzle in two different
ways. First, after generating the rotating and accelerating Kerr-NUT metric via the inverse scattering
method (ISM) \cite{JETP48-985,JETP50-1,BV2005} from the seed Rindler vacuum background, we obtain
two distinct forms of the same solution by using two different parametrizations. This clearly shows
that both metrics are isometric to each other. Then, we find the obvious M\"{o}bius transformation
and parameter identifications between them. In the meanwhile, we will illustrate what limit should
be taken in the general P-D solution to obtain the type-D NUT C-metric. In particular, we shall
present another M\"{o}bius transformation and linear combinations of the Killing coordinates
that clearly cast the type-D NUT C-metric into the familiar form of the P-D solution, with
the quartic structure function being factorized into two quadratic functions in a manner different
from that of Griffiths-Podolsk\'{y}.

The main parts of our paper are organized as follows. In Sec. \ref{KerrNUT}, we present the ISM
construction of the type-D accelerating Kerr-NUT metric as a 2-soliton transformation from the
Rindler vacuum background, and two distinct parametrizations to arrive at two different rational
expressions of the same solution. As preparation, we take the C-metric as a simple example to
compare the double Kerr-Schild formalism with the ISM perspective, and make an attempt to combine
these two seemly unrelated schemes. This idea has never been advocated in the past years. Then
after a minor retrospect of the ISM algorithm, we devise a new strategy for the normalization of
the obtained metric derived via the ISM construction from the vacuum seed solution, which could
facilitate the construction of higher dimensional solutions by using the trivial vacuum background
as the seed metric. Our normalization scheme is corroborated in the Appendix \ref{Appe} by deriving
the five-dimensional Schwarzschild-Tangherlini solution \cite{NC27-636} from the flat vacuum background
in two different ways. The remaining two sections, \ref{Mtct} and \ref{NUTCPD}, are entirely new and
constitute the main contribution of this work, where the M\"{o}bious transformation of the radial and
angular coordinates, and linear combinations of the Killing coordinates as well as various parameter
identifications between these two metrics are plainly displayed. Our paper ends with a brief summary
of this work, followed by the Aappendix with two simple examples to illuminate our new normalization
strategy.

\section{Accelerating Kerr-NUT solution: ISM construction
and two different parametrizations}\label{KerrNUT}

\subsection{The C-metric and its seed}\label{ssCm}

We begin with two different descriptions of the C-metric, whose line element in the spherical-like
coordinates ($t, r, x=\cos\theta, \bp$),
\be\label{Cmds}
d\ts^2 = \frac{1}{(1 +\alpha\,rx)^2}\Big[-f(r)dt^2 +\frac{dr^2}{f(r)}
 +r^2\frac{dx^2}{h(x)} +r^2h(x)d\tp^2 \Big] \, ,
\ee
can be rewritten in terms of the soliton formalism as
\be\label{Cm}
d\ts^2 \equiv \alpha\, ds^2 = -\frac{\mu_1\mu_3}{\mu_2}dt^2 +\frac{\rho^2\mu_2}{\mu_1\mu_3}d\phi^2
 +\frac{\mu_1\mu_3R_{12}^2R_{23}^2}{(1-2m\alpha)^2\mu_2R_{13}^2R_{11}R_{22}R_{33}}
 \big(d\rho^2 +dz^2\big) \, ,
\ee
where the Weyl coordinates are defined as
\bea
\rho = \frac{\sqrt{r^2f(r)h(x)}}{(1 +\alpha\,rx)^2} \, , \quad
z = \frac{(\alpha\, r +x)[(1+m\alpha\,x)r -m]}{(1 +\alpha\,rx)^2} \, , \nn
\eea
with two structure functions being
\be
f(r) = \big(1 -\alpha^2r^2\big)\Big(1 -\frac{2m}{r}\Big) \, , \qquad
h(x) = \big(1 -x^2\big)(1 +2m\alpha\,x) \, .
\ee
In the above, we have also made a rescaling $\tp = \alpha\phi$ and introduced the following notations:
\be
R_{kl} = \rho^2 +\mu_k\mu_l \, , \quad \mu_k = \sqrt{\rho^2 +(z -z_k)^2} -(z -z_k) \, ,
\ee
where the 3-soliton $\mu_k$s are algebraically expressed as
\be\label{3Cs}
\mu_1 = \frac{(1 -\alpha\, r)(r -2m)(1-x)}{(1 +\alpha\,rx)^2} \, , \quad
\mu_2 = \frac{(1 -\alpha\, r)r(1-x)(1 +2m\alpha\, x)}{(1 +\alpha\,rx)^2} \, , \quad
\mu_3 = \frac{(1 -\alpha^2r^2)(1 +2m\alpha\, x)}{\alpha(1 +\alpha\,rx)^2} \, ,
\ee
which are associated with three rod end points: $z_1 = -m \, , z_2 = m \, , z_3 = 1/(2\alpha)$,
respectively.

The zero-mass limit $\mu_2\mapsto \mu_1$ is the Rindler vacuum
\be\label{seed}
d\bs^2 \equiv \alpha\, ds^2 = -\mu_3 dt^2 +\frac{\rho^2}{\mu_3}d\phi^2
 +\frac{\mu_3}{\rho^2 +\mu_3^2}\big(d\rho^2 +dz^2\big) \, ,
\ee
which will be used later as our seed metric. The Rindler metric can be viewed as 1-soliton
transformation from the flat Minkowski background with the metric components: $g_{\mu\nu} = (-1,
\rho^2, 1, 1)$. However, it is noted that the Rindler vacuum is trivial and is entirely equivalent
to the flat Minkowski metric since they are related via some simple coordinate transformations.

So in the soliton formalism, the C-metric can be interpreted either as a 3-soliton transformation
to the flat Minkowski background or a 2-soliton transformation to the Rindler background. In this
article, we will take the latter viewpoint and derive the accelerating Kerr-NUT solution via the
ISM procedure by using the Rindler background (\ref{seed}) as the seed metric. Note that usually
the seed solution need not be exactly the background metric and frequently can be taken as a
more complicated even singular one for the technical aim. However, a simple background metric
is surely more preferable in practice as a seed solution.

On the other hand, the C-metric (\ref{Cmds}) admits the double Kerr-Schild formulation as follows:
\bea
d\ts^2 &=& \frac{1}{(1 +\alpha\,rx)^2}\Big[-\big(1 -\alpha^2r^2\big)d\bt^2
 +\frac{dr^2}{1 -\alpha^2r^2} +r^2\frac{dx^2}{1-x^2} +r^2\big(1 -x^2\big)d\bp^2 \Big] \nn \\
&& +\frac{2m\big(1 -\alpha^2r^2\big)}{r(1 +\alpha\,rx)^2}\Big(d\bt +\frac{dr}{1 -\alpha^2r^2}\Big)^2
 +\frac{2m\alpha\,x\big(1 -x^2\big)r^2}{(1 +\alpha\,rx)^2}\Big(d\bp +\frac{I\, dx}{1-x^2}\Big)^2 \, ,
\eea
after making the coordinate transformations
\bea
t = \bt -\int\frac{2m}{rf(r)}dr \, , \quad
\tp = \bp +I\int\frac{2m\alpha\,x}{h(x)}dx \, . \nn
\eea
The zero-mass part is just the Rindler vacuum written in the C-metric coordinates ($x, y$) with
$y = -1/(\alpha\, r)$. In view of this, the C-metric can be interpreted as a perturbation around
the Rindler background. Incidentally, it should be pointed out that the above Kerr-Schild ansatz
is completely applicable to the most general P-D metric \cite{AP98-98}.

However, it seems that the following double quasi-Kerr-Schild prescription,
\bea
d\ts^2 &=& \frac{1}{(1 +\alpha\,rx)^2}\Big[-\big(1 -\alpha^2r^2\big)dt^2
 +\frac{dr^2}{f(r)} +r^2\frac{dx^2}{h(x)} +r^2\big(1 -x^2\big)d\tp^2 \Big] \nn \\
&& +\frac{2m\big(1 -\alpha^2r^2\big)}{r(1 +\alpha\,rx)^2}dt^2
 +\frac{2m\alpha\,x\big(1 -x^2\big)r^2}{(1 +\alpha\,rx)^2}d\tp^2 \, ,
\eea
is in more accordance with the above 2-soliton viewpoint by virtue of the two-dimensional metrics
are conformal to each other:
\bea
\frac{1}{(1 +\alpha\,rx)^2}\Big[\frac{dr^2}{f(r)} +r^2\frac{dx^2}{h(x)}\Big]
 \simeq C_f(\rho, z)(d\rho^2 +dz^2) \, . \nn
\eea

\subsection{ISM: A minor preliminary}

In a $D$-dimensional spacetime with $(D-2)$ Killing vectors, if its metric, whose components
depend on the Weyl coordinates $\rho$ and $z$ only, is written as \cite{PRD70-124002}
\be
ds^2 = g_{ab} dx^a dx^b +f(\rho, z)(d\rho^2+dz^2) \, ,
\ee
and supplemented with the canonical condition,
\be\label{cc}
\det \bg = -\rho^2 \, ,
\ee
then the vacuum Einstein equations are equivalent to the following equations for the $(D-2)\times
(D-2)$ matrix $\bg = (g_{ab})$:
\be\label{UV}
\pd_{\rho}\bU +\pd_z\bV = 0 \, ,
\ee
and for the conformal factor $f(\rho,z)$, which can be integrated by quadratures:
\be\begin{split}
\pd_\rho \ln\, f &= -\frac{1}{\rho} +\frac{1}{4\rho}\Tr(\bU^2 -\bV^2) \, , \\
\pd_z \ln\, f &= \frac{1}{2\rho}\Tr (\bU\bV) \, ,
\end{split}\ee
where the following two matrices are introduced
\be
\bU = \sqrt{-g}(\pd_\rho\bg)\bg^{-1} \, ,\qquad
\bV = \sqrt{-g} (\pd_z\bg)\bg^{-1} \, .
\ee

System (\ref{UV}) is completely integrable and can be viewed as the compatibility condition
of the Lax pair for the $(D-2)\times(D-2)$ generating matrix $\bPsi(\lambda,\rho,z)$:
\be\begin{split}
& \pd_{\rho}\bPsi +\frac{2\lambda\rho}{\lambda^2 +\rho^2}\pd_{\lambda}\bPsi =
 \frac{\rho\bU +\lambda\bV}{\lambda^2 +\rho^2}\bPsi \, , \\
& \pd_z\bPsi -\frac{2\lambda^2}{\lambda^2 +\rho^2}\pd_{\lambda}\bPsi =
 \frac{\rho\bV -\lambda\bU}{\lambda^2 +\rho^2}\bPsi \, ,
\end{split}\ee
with $\lambda$ being a complex spectral parameter independent of $\rho$ and $z$. When $\lambda = 0$,
one gets $\bg(\rho,z) = \bPsi(0,\rho,z)$. One can construct new solitonic solutions from a seed
metric via the dressing procedure:
\be
\bPsi = \bichi\, \bPsi_0 \, , \qquad
\bichi = \mathbf{I} +\sum_{k=1}^n \frac{\bn^{\dag(k)}\bm^{(k)}}{\lambda-\mu_k} \, ,
\ee
in which $\bn^{\dag(k)}$ denotes the transpose of the BZ vector $\bn^{(k)}$ defined below, $\mu_k$
represents a soliton with its corresponding antisoliton being $\nu_k = -\rho^2/\mu_k$, and the
constant $z_k$ is the turning point (or rod end point) on the $z$ axes.

The expression for the generated matrix of the general $n$-soliton solutions is
\be
\widetilde{\bg} = \bg_0 -\sum_{k,l=1}^n
 \frac{\bl^{\dag(k)}(\bGa^{-1})_{kl}\bl^{(l)}}{\mu_k\mu_l} \, ,
\ee
with the following notations being introduced
\be
\bGa_{kl} = \frac{\bm^{(k)} \bg_0 \bm^{\dag(l)}}{\rho^2 +\mu_k\mu_l} \, , \qquad
 \bl^{(k)} = \bm^{(k)} \bg_0 \, , \qquad
 \bm^{(k)} = \bm_0^{(k)} \bPsi_0^{-1}(\mu_k,\rho,z) \, , \qquad
 \bn^{(k)} = \sum_{l=1}^n\frac{(\bGa^{-1})_{kl}\bl^{(l)}}{\mu_l} \, ,
\ee
where the BZ vectors $\bm_0^{(i)}$ consist of arbitrary constants with one of which can be set
to unity for each vector.

In the case of a diagonal seed, the generating matrix $\bPsi_0(\mu_i,\rho,z)$ can be easily obtained
from the seed metric $\bg_0(\rho,z)$ by simply replacing every soliton or antisoliton via $\mu_k\to
\mu_k -\lambda$ and $\rho^2\to \rho^2 -2\lambda\, z -\lambda^2$, or $\rho^2/\mu_k\to \rho^2/\mu_k
+\lambda$.

In general, the generated metric $\widetilde{\bg}$ is unphysics due to the fact that it does not
fulfill the canonical condition (\ref{cc}): $\det \tilde{\bg} \not= -\rho^2$, so it should be
appropriately normalized.

As far as the normalization scheme is concerned, to the best of our knowledge, there exist three
different kinds of strategies to date. (I) Uniform normalization \cite{BV2005}: Like the
four-dimensional case, all the metric components of the Killing parts in the higher-dimensional
case are multiplied by the same normalization factor. The main defect of this method lies in that
it yields some fractional singularities on the symmetry axes, so that the final solution has little
physics meaning. (II) Partial block uniform normalization \cite{PRD73-064009,PRD73-124034,PRD74-104004,
PTP116-319}, which stems from early work \cite{PRD34-2990,PLA136-269}: This method only applies the
above uniform normalization to a part of the block diagonal matrix of the generated metric, so it is
only effective in the case when not all rotation parameters are turned on, for instance, it is limited
to the singly rotating case in five dimensions, and so on. (III) Pomeransky's \emph{remove--re-add}
tricky \cite{PRD73-044004}: If the seed metric already satisfies the canonical condition, then
after removing and re-adding the same number of solitions or antisolitons, the generated metric
automatically subjects to the canonical condition also. At the same time, the conformal factor can
be simply computed as $f = f_0\det\bGa/{\det\bGa_0}$, where $\bGa_0$ can be obtained from $\bGa$
by setting the nontrivial BZ constants to zero. This is a wise strategy that is widely adopted
nowadays, for example, the doubly rotating Myers-Perry solution \cite{AP172-304} is regenerated
via this method from the five-dimensional Schwarzschild-Tangherlini solution \cite{NC27-636}.
However, there is a fly in the ointment in that it cannot work effectively in the case when the
seed solution is a vacuum background metric where there is no soliton that can be removed. In
particular, it is claimed \cite{PRD73-044004} that the five-dimensional Schwarzschild-Tangherlini
solution \cite{NC27-636} cannot be obtained as a 2-soliton solution on the flat Minkowski background.

Below we propose the fourth scheme for the normalization of the obtained metric derived from the
vacuum seed solution, which would be potentially useful for the construction of higher dimensional
solutions by using the trivial vacuum background as the seed metric. In the Appendix, we will
present two simple examples to illuminate this method. First, let us consider a simple ansatz:
\be
\bg = \bW\widetilde{\bg}\bW \, , \qquad  \bW = \diag(w_1 \, , \cdots \, , w_{D-2}) \, ,
\ee
by simply employing a diagonal matrix $\bW$ with which each of its entries satisfies the Laplace
equation:
\be\label{Laeq}
\nabla^2\ln\, w_i = \Big(\pd_\rho^2 +\frac{1}{\rho}\pd_\rho +\pd_z^2 \Big)\ln\, w_i = 0 \, .
\ee
The solution of each entry in the above equation has the general form: $w_i = \rho^{b_i}\mu_{i}^{c_i}$,
in which $b_i$ and $c_i$ are arbitrary constants. The entries in $\bW$ are chosen so that $-\rho^2
= w_1^2\cdots\,w_{D-2}^2\det\widetilde{\bg}$ is satisfied. By virtue of the fact that $\det
\widetilde{\bg}$ is proportional to $\det\bg_0$, actually one only needs to choose different
powers in $w_i$.

If the generated metric is still diagonal just as its seed metric, then the above means amounts to
$\bg = \bW^2\widetilde{\bg}$, and it is easy to compute the conformal factor. On the other hand,
if all the diagonal entries are identical $w_i = w$, then the above scheme is equivalent to the
original uniform normalization strategy. Suppose that only some of the diagonal entries are set
to be identical, then this amounts to the partial block uniform normalization. Finally, if all
$w_i$s are appropriately chosen such as ($-\rho^2/\mu_{i}^2$) or ($-\mu_{i}^2/\rho^2$), this
scheme is actually equivalent to removing solition or antisoliton at least in the case of the
generated metric being static (diagonal).

Second, let us handle the more general rotating case. Since the generated metric in general still needs to
make further linear combinations of the Killing coordinates, that is, $\hat{\bg} = \bA^{\dag}\bg\bA$,
where the rotation matrix $\bA$ is unimodular and its every entry is a constant, then we can have
$\hat{\bg} = \hat{\bW}^{\dag}\widetilde{\bg}\hat{\bW}$, where $\hat{\bW} = \bW\bA$, $\hat{\bW}^{\dag}$
is the matrix transpose of $\hat{\bW}$ and each of its entry obeys the above Laplace equation
(\ref{Laeq}). Clearly it is troublesome to choose a suitable matrix $\hat{\bW}$ so that the final
metric is the expected one. However, one need not worry about this, since the Pomeransky's
\emph{remove--re-add} tricky is already very effective to generate the rotating solution, so our
method just plays a nice complement to the case where the Pomeransky's tricky fails to work.

\subsection{ISM construction of
the accelerating Kerr-NUT solution}

Although the accelerating Kerr-NUT solution can be derived as a 3-soliton transformation via the
ISM procedure from the flat Minkowski background, in accordance with the above double Kerr-Schild
perspective, we prefer to rederive it as a 2-soliton transformation by using the Rindler background
(\ref{seed}) as the seed metric, whose Killing part is
\be\label{Rindler}
\bg_0 = \diag(-\mu_3\, , \rho^2/\mu_3) = \diag(-\mu_3\, , -\nu_3) \, ,
\ee
with the conformal factor $f_0 = \mu_3/R_{33}$, so that the generating matrix is easily obtained as
\be
\bPsi_0 = \diag(\lambda -\mu_3\, , \lambda -\nu_3)
 = \diag(\lambda -\mu_3\, , \lambda +\rho^2/\mu_3) \, .
\ee
Now we perform a 2-soliton transformation to this Rindler background: add a soliton $\mu_1$ at
$z = z_1$ with the vector $m_0^{(1)} = (C_1\, , 1)$ and a soliton $\mu_2$ at $z = z_2$ with the
vector $m_0^{(2)} = (1\, , C_2)$, respectively. Note that we take the 2-soliton in the order:
$\mu = (\mu_1\, , \mu_2)$, so that the generated metric exactly reproduces the same form (\ref{Cm})
for the C-metric when the BZ parameters are set to zero: $C_1 = C_2 =0$. Then two BZ vectors
constitute a $2\times\, 2$ matrix:
\be
\bm = \bigg(\begin{array}{cc}
 \frac{C_1}{\mu_{13}} & \frac{\mu_3}{R_{13}} \\
 \frac{1}{\mu_{23}} & \frac{C_2\mu_3}{R_{23}}
\end{array} \bigg) \, , \quad
\bl = \bigg(\begin{array}{cc}
 \frac{-C_1\mu_3}{\mu_{13}} & \frac{\rho^2}{R_{13}} \\
 \frac{-\mu_3}{\mu_{23}} & \frac{C_2\rho^2}{R_{23}}
\end{array} \bigg)
\ee
where a short notation is included for briefness: $\mu_{ij} = \mu_i -\mu_j$. The $\bGa$ matrix
is also recorded as follows:
\be
\bGa = \bigg(\begin{array}{cc}
\frac{-C_1^2\mu_3}{\mu_{13}^2R_{11}} +\frac{\mu_3\rho^2}{R_{13}^2R_{11}}
 & \frac{-C_1\mu_3}{\mu_{13}\mu_{23}R_{12}} +\frac{C_2\mu_3\rho^2}{R_{12}R_{13}R_{23}} \\
\frac{-C_1\mu_3}{\mu_{13}\mu_{23}R_{12}} +\frac{C_2\mu_3\rho^2}{R_{12}R_{13}R_{23}}
 & \frac{-\mu_3}{\mu_{23}^2R_{22}} +\frac{C_2^2\mu_3\rho^2}{R_{23}^2R_{22}}
\end{array} \bigg) \, .
\ee

The final physics metric can be obtained via either the uniform normalization scheme or our
prescription proposed in the above
\be
\bg = \bW\widetilde{\bg}\bW = -\frac{\mu_1\mu_2}{\rho^2}\widetilde{\bg} \, ,
\ee
in which we have taken $\bW = \diag(\sqrt{-\mu_1\mu_2}/\rho \, , \sqrt{-\mu_1\mu_2}/\rho)$.
The conformal factor is
\be
f = -16k^2f_0\frac{\mu_1^3\mu_2^3}{\rho^2\mu_{12}^2}\det\bGa = 16k^2\frac{B}{H} \, .
\ee

Then the generated metric represents the accelerating Kerr-NUT solution, whose line element reads
\be\label{acKNweyl}
ds^2 = \frac{-\mu_3A}{\mu_1\mu_2B} \Big(dt +\frac{\mu_{12}C}{\mu_3A}d\phi\Big)^2
 +\frac{\rho^2\mu_1\mu_2B}{\mu_3A}d\phi^2 +16k^2\frac{B}{H}\big(d\rho^2 +dz^2\big) \, ,
\ee
where
\bea
A &=& \big[C_1(C_2\mu_2\mu_{23}R_{12} -\rho\mu_{12}R_{23})R_{13}
 -\mu_1\mu_{13}(C_2\rho\mu_2\mu_{12}\mu_{23} +R_{12}R_{23})\big] \nn \\
&& \times\big[C_1(C_2\mu_2\mu_{23}R_{12} +\rho\mu_{12}R_{23})R_{13}
 +\mu_1\mu_{13}(C_2\rho\mu_2\mu_{12}\mu_{23} -R_{12}R_{23})\big] \, , \nn \\
B &=& C_1^2\big(C_2^2\mu_{23}^2R_{12}^2 +\mu_{12}^2R_{23}^2\big)R_{13}^2
 -2C_1C_2\mu_{13}\mu_{23}R_{13}R_{23}R_{11}R_{22} \nn \\
&& +\big(C_2^2\rho^4\mu_{12}^2\mu_{23}^2 +R_{12}^2R_{23}^2\big)\mu_{13}^2 \, , \nn \\
C &=& C_1\mu_2\mu_{13}\big(C_2^2\rho^2\mu_{23} -R_{23}^2\big)R_{11}R_{12}R_{13}
 +C_2\mu_1\mu_{23}\big(C_1^2R_{13}^2 -\rho^2\mu_{13}^2\big)R_{12}R_{22}R_{23} \, , \nn \\
H &=& \frac{\mu_{12}^2\mu_{13}^2\mu_{23}^2}{\mu_1^3\mu_2^3\mu_3^3}R_{12}^2R_{13}^2R_{23}^2
 R_{11}R_{22}R_{33} \, . \nn
\eea
When $C_1 = C_2 = 0$, the above generated 2-soliton solution just reduces to the C-metric (\ref{Cm}).
So according to the perspective in Sec. \ref{ssCm}, the generated accelerating Kerr-NUT
solution (\ref{acKNweyl}) can be interpreted as a Kerr-NUT black hole sitting on the Rindler
background.

Note that the above accelerating Kerr-NUT solution (\ref{acKNweyl}) can also be regenerated via the
\emph{``remove--re-add"} tricky from the C-metric (\ref{Cm}). The process is briefly sketched as 
follows. We first remove an antisoliton $\nu_1$ at $z = z_1$ and a soliton $\mu_2$ at $z = z_2$ 
both with the trivial vectors $(1, 0)$, then rescale the metric by a factor $\mu_1/\mu_2$. This 
essentially generates the above Rindler seed (\ref{Rindler}):
\bea
\bg_0 = \frac{\mu_1}{\mu_2}\diag\Big(\frac{\mu_2^2}{\mu_1^2}\, , 1\Big)\times
\diag\Big(-\frac{\mu_1\mu_3}{\mu_2}\, , \frac{\rho^2\mu_2}{\mu_1\mu_3}\Big) \, . \nn
\eea
Next we re-add the antisoliton $\nu_1$ with the vector $m_0^{(1)} = (1, -C_1$) and the soliton
$\mu_2$ with the vector $m_0^{(2)} = (1, C_2$), and multiply the generated metric by a factor
$\mu_2/\mu_1$ so that the final metric is physics because it satisfies the canonical condition:
$\det\bg = -\rho^2$. The conformal factor can be easily evaluated via $f = f_{00}\det\bGa/{\det
\bGa(C_1=0, C_2=0)}$, where $f_{00}$ is the conformal factor given by Eq. (\ref{Cm}).

\subsection{Astorino's parametrization}

We first transform from the Weyl-Lewis-Papapetrou coordinates ($\rho, z$) to the spherical-like
coordinates ($r, x$) via
\be
\rho = \frac{\sqrt{\Delta(r)P(x)}}{(1 +\alpha\, rx)^2} \, , \qquad
z = \frac{(x +\alpha\, r)\big[(r-m)(1 +m\alpha\, x)
 +\sigma^2\alpha\, x\big]}{(1 +\alpha\, rx)^2} \, , \nn
\ee
where
\bea
\Delta(r) = \big[(r-m)^2 -\sigma^2\big]\big(1 -\alpha^2r^2\big) \, , \quad
P(x) = \big(1 -x^2\big)\big[(1 +m\alpha\, x)^2 -\sigma^2\alpha^2x^2\big] \, ,
\eea
in which $\sigma = \sqrt{m^2 +n^2 -a^2}$.

The locations of three turning points are taken to be
\bea
z_1 = -\sigma, \quad z_2 = \sigma, \quad z_3 = \frac{1 -\alpha^2(m^2 -\sigma^2)}{2\alpha} \, , \nn
\eea
so their corresponding solitons are purely algebraic:
\bea
\mu_1 &=& \frac{(1 -\alpha\, r)(r -m -\sigma)[1
+(m -\sigma)\alpha\, x)](1-x)}{(1 +\alpha\, rx)^2} \, , \nn \\
\mu_2 &=& \frac{(1 -\alpha\, r)(r -m +\sigma)[1
+(m +\sigma)\alpha\, x)](1-x)}{(1 +\alpha\, rx)^2} \, , \\
\mu_3 &=& \frac{\big(1 -\alpha^2r^2\big)[1
+(m -\sigma)\alpha\, x)][1 +(m +\sigma)\alpha\, x)]}{\alpha
(1 +\alpha\, rx)^2} \, , \nn
\eea
which reduce to Eq. (\ref{3Cs}) exactly when the static C-metric limit is reached ($\sigma = m$).

It is very tedious to convert the four functions ($A, B, C, H$) into the expressions in terms
of the spherical-like coordinates. However, if we take
\be
C_1 = \frac{(m-\sigma)[1 +(m +\sigma)\alpha]}{(a-n)[1 -(m -\sigma)\alpha]} \, , \quad
C_2 = \frac{(m-\sigma)[1 -(m +\sigma)\alpha]}{(a+n)[1 +(m -\sigma)\alpha]} \, ,
\ee
and
\be
k = \frac{(m+\sigma)[1 +(m -\sigma)\alpha][1
 -(m -\sigma)\alpha]}{2\alpha^{3/2}\sqrt{1 +\alpha^2a^2}} \, ,
\ee
then we are able to first write the two-dimensional metric as
\bea
f\big(d\rho^2 +dz^2\big) = \frac{\Sigma}{\big(1 +\alpha^2a^2\big)
(1 +\alpha\,rx)^2}\Big[\frac{dr^2}{\Delta(r)} +\frac{dx^2}{P(x)}\Big] \, , \nn
\eea
in which
\be
\Sigma = r^2 +(ax -n)^2 +2n\alpha\,r\big[a\big(1+x^2\big)-2nx\big] +\alpha^2r^2(a-nx)^2
 +\big(a^2-n^2\big)^2\alpha^2x^2 \, .
\ee
Finally, after some cumbersome algebraic simplifications, the Killing parts can be converted into
\bea
g_{ab} dx^a dx^b = \frac{-F}{(1 +\alpha\,rx)^2\Sigma\alpha}\Big(dt -\frac{\omega\alpha}{F}
 d\varphi\Big)^2 +\frac{\Sigma\Delta(r)P(x)\alpha}{(1 +\alpha\,rx)^2F}d\varphi^2 \, , \nn
\eea
where
\bea
F &=& \big[1 -\big(a^2-n^2\big)\alpha^2x^2\big]^2\Delta(r)
 -\big[a\big(1+\alpha^2r^2\big) +2n\alpha\,r\big]^2P(x) \, , \nn \\
\omega &=& \big[a\big(1+x^2\big)-2nx\big]\big[1 -\big(a^2-n^2\big)\alpha^2x^2\big]\Delta(r)
 +\big(r^2+n^2-a^2\big)\big[a\big(1+\alpha^2r^2\big) +2n\alpha\,r\big]P(x) \, . \nn
\eea

After rescaling the Killing coordinates as $t\mapsto \sqrt{\alpha}t\, , \varphi\mapsto
\varphi/\sqrt{\alpha}$, the above two parts are collectively written in terms of the
Lewis-Papapetrou coordinates as \cite{PRD109-084038}
\be\label{AcKN}
ds^2 = \frac{-F}{(1 +\alpha\,rx)^2\Sigma}\Big(dt -\frac{\omega}{F}d\varphi\Big)^2
 +\frac{\Sigma}{(1 +\alpha\,rx)^2F}\bigg\{\Delta(r)P(x)d\varphi^2
 +\frac{F}{1+\alpha^2a^2}\Big[\frac{dr^2}{\Delta(r)} +\frac{dx^2}{P(x)}\Big]\bigg\} \, ,
\ee
which can also be recast into the Boyer-Lindquist-like form
\be\label{Ast}
ds^2 = \frac{1}{(1 +\alpha\,rx)^2}\bigg\{\frac{-\Delta(r)X^2 +P(x)Y^2}{\Sigma}
 +\frac{\Sigma}{1+\alpha^2a^2}\Big[\frac{dr^2}{\Delta(r)} +\frac{dx^2}{P(x)}\Big]\bigg\} \, ,
\ee
where
\be\begin{split}
X &= \big[1 -\big(a^2-n^2\big)\alpha^2x^2\big]dt
  +\big[2nx -a\big(1+x^2\big)\big]d\varphi \, , \\
Y &= \big[a\big(1+\alpha^2r^2\big) +2n\alpha\,r\big]dt
  +\big(r^2+n^2-a^2\big)d\varphi \, .
\end{split}\ee

In Sec. III (especially Sec. IIIB) of Ref. \cite{PRD109-084038}, the above metric
(\ref{AcKN}) was derived through an equivalent limiting procedure ($a_2\to 0, l_2\to 0$ and
then $w_4\to\infty$) on the double Kerr-NUT solution. It was also claimed there that this
spacetime is of type D, and might be included into the P-D family of black holes. Here we
have built this metric with two solitons on a Rindler accelerating background. In the subsequent
subsections, we will give an explicit verification of the Astorino's insight by showing that the
solution (\ref{Ast}) is indeed diffeomorphic to the P-D metric.

In the absence of the NUT charge, the above metric simply recovers the accelerating Kerr solution
in the familiar form after further shifting the Killing coordinates:
\bea
t\mapsto \frac{t -2a\varphi}{\sqrt{1+\alpha^2a^2}} \, , \quad
\varphi\mapsto -\frac{\alpha^2a t +\big(1 -\alpha^2a^2\big)\varphi}{\sqrt{1+\alpha^2a^2}} \, . \nn
\eea
On the other hand, the accelerating NUT C-metric takes a simple expression:
\be\label{NUTC}
ds^2 = \frac{1}{(1 +\alpha\,rx)^2}\bigg\{ -\frac{\Delta(r)}{\Sigma}\big[\big(1 +n^2\alpha^2x^2\big)dt
 +2nx d\varphi\big]^2 +\frac{P(x)}{\Sigma}\big[2n\alpha\,r dt +\big(r^2+n^2\big)d\varphi\big]^2
 +\Sigma\Big[\frac{dr^2}{\Delta(r)} +\frac{dx^2}{P(x)}\Big]\bigg\} \, ,
\ee
in which
\be\begin{split}
\Delta(r) &= \big(r^2 -2mr -n^2\big)\big(1 -\alpha^2r^2\big) \, , \quad
P(x) = \big(1 -x^2\big)\big(1 +2m\alpha\, x -n^2\alpha^2x^2\big) \, , \\
\Sigma &= \big(r^2 +n^2\big)\big(1 +n^2\alpha^2x^2\big) -4n^2\alpha\, rx \, .
\end{split}\ee
When the accelerating parameter is set to zero, the above type-D NUT C-metric apparently reduces
to the Taub-NUT solution, and also it obviously recovers the usual C-metric when the NUT charge
vanishes. There is a natural question remaining, namely, whether it is superficially accelerating
just as pointed out in Ref. \cite{CQG22-3467} for the traditional form. Then, if it is so, we
still lack a type-D NUT C-metric. However, this case seems unlikely. So, we will assume that
the above NUT C-metric is not isometric to the Taub-NUT solution in the remaining context, but
clearly an obvious proof is still needed.

The above solution has also been generalized in Ref. \cite{2404.06551} to include the electric
and magnetic charges as well as a nonzero cosmological constant. However, it suffices enough for
our aim to turn off these three parameters. Also, in our opinion, it becomes useful only in the
static case (where the rotation parameter vanishes), as we will demonstrate the above rotating
and accelerating solution can be related to the traditional form of the general P-D metric.

\subsection{Griffiths-Podolsk\'{y}'s parametrization}

We next transform from the Weyl-Lewis-Papapetrou coordinates ($\rho, z$) to the spherical-like
coordinates ($\tr, \tx$) via
\be
\rho = \frac{\sqrt{\tD(\tr)\tP(\tx)}}{[1 +\talpha\tr(\tn+\ta\tx)]^2} \, , \qquad
z = \frac{\big[\tx +\talpha\tr(\ta +\tn\tx)\big]\big[\tr -\tm +\talpha(\tn +\ta\tx)
 \big(\tm\tr +\tn^2 -\ta^2\big)\big]}{\big[1 +\talpha\tr(\tn +\ta\tx)\big]^2} \, , \nn
\ee
where
\be\begin{split}
\tD(\tr) &= \big(\tr^2 -2\tm\tr +\ta^2 -\tn^2\big)\big[(1 +\talpha\tn\tr)^2
 -\talpha^2\ta^2\tr^2\big] \, , \\
\tP(\tx) &= \big(1 -\tx^2\big)\big[1 +2\tm\talpha(\tn +\ta\tx)
 +\big(\ta^2 -\tn^2\big)\talpha^2(\tn +\ta\tx)^2\big] \, ,
\end{split}\ee
in which $\tsigma = \sqrt{\tm^2 +\tn^2 -\ta^2}$.

The locations of three rod end points are now taken to be
\bea
z_1 = -\tsigma, \quad z_2 = \tsigma, \quad z_3 = \frac{1 +2\talpha\tm\tn
 -\talpha^2\big(\ta^2-\tn^2\big)^2}{2\talpha\ta} \, , \nn
\eea
and their corresponding solitons are also purely algebraic:
\bea
\mu_1 &=& \frac{[1 -\talpha(\ta-\tn)\tr](\tr -\tm -\tsigma)[1
 +(\tm -\tsigma)\talpha(\tn+\ta\tx)](1-\tx)}{[1 +\talpha\tr(\tn+\ta\tx)]^2} \, , \nn \\
\mu_2 &=& \frac{[1 -\talpha(\ta-\tn)\tr](\tr -\tm +\tsigma)[1
 +(\tm +\tsigma)\talpha(\tn+\ta\tx)](1-\tx)}{[1 +\talpha\tr(\tn+\ta\tx)]^2} \, , \\
\mu_3 &=& \frac{[1 +\talpha(\ta+\tn)\tr][1 -\talpha(\ta-\tn)\tr][1
 +(\tm -\tsigma)\talpha(\tn+\ta\tx)][1 +(\tm +\sigma)\talpha(\tn+\ta\tx)]}{\talpha
[1 +\talpha\tr(\tn+\ta\tx)]^2} \, . \nn
\eea

It is troublesome to convert the four functions ($A, B, C, H$) into the expressions in terms
of the spherical-like coordinates. A simple trick to do this relatively easily is to avoid using
the square root, such as, exploiting identities: $\tsigma^{2k} = \big(\tm^2 +\tn^2 -\ta^2\big)^k$
and $\tsigma^{2k+1} = \big(\tm^2 +\tn^2 -\ta^2\big)^k\tsigma$ can efficiently facilitate to
rationalize the expressions containing the square root. This experience is also very effective
to manipulate the phantom soliton whose square-root function could not be reduced to a simple
algebraic expression.

In order to let the conformal factor include a familiar multiplier $\tSigma = \tr^2 +(\tn
+\ta\tx)^2$, we have to take
\be
C_1 = \frac{(\ta-\tn)[1 +(\tm +\tsigma)\talpha(\ta+\tn)]}{(\tm+\tsigma)
 [1 -(\tm -\tsigma)\talpha(\ta-\tn)]} \, , \quad
C_2 = \frac{(\ta+\tn)[1 -(\tm +\tsigma)\talpha(\ta-\tn)]}{(\tm+\tsigma)
 [1 +(\tm -\tsigma)\talpha(\ta+\tn)]} \, ,
\ee
and
\be
k = \frac{(\tm+\tsigma)[1 +(\tm -\tsigma)\talpha(\ta+\tn)][1 -(\tm -\tsigma)
 \talpha(\ta-\tn)]}{2(\talpha\ta)^{3/2}\sqrt{1 +\talpha^2\big(\ta^2-\tn^2\big)^2}} \, .
\ee

Finishing the remaining task of the algebraic process, we find that it is instructive to cast
the metric into the Boyer-Lindquist-like form rather than the Lewis-Papapetrou form, which is
rather concise and simple enough:
\bea\label{PD0}
d\ts^2 = \frac{1}{\big[1 +\talpha\tr(\tn +\ta\tx)\big]^2}\bigg\{\frac{-\tD(\tr)\tX^2
 +\tP(\tx)\tY^2}{\tSigma} +\tSigma\Big[\frac{d\tr^2}{\tD(\tr)}
 +\frac{d\tx^2}{\tP(\tx)}\Big]\bigg\} \, ,
\eea
where $\tSigma = \tr^2 +(\tn +\ta\tx)^2$ and
\be\begin{split}\label{tXtY}
\tX &= \frac{\big[1 -\big(\ta^2-\tn^2\big)\talpha^2(\tn +\ta\tx)^2\big]dt
 +\talpha\ta\big[2\tn\tx +\ta\big(1+\tx^2\big)\big]d\varphi}{\sqrt{\talpha\ta}
 \sqrt{1 +\talpha^2\big(\ta^2-\tn^2\big)^2}} \, , \\
\tY &= \frac{\ta\big[1+\big(\ta^2-\tn^2\big)\talpha^2\tr^2\big]dt
  -\talpha\ta\big(\tr^2+\tn^2-\ta^2\big)d\varphi}{\sqrt{\talpha\ta}
 \sqrt{1 +\talpha^2\big(\ta^2-\tn^2\big)^2}} \, .
\end{split}\ee
In the above frame, the Rindler background or the accelerating horizons is rest relative to the
conformal infinity.

One can note that the above metric resembles somewhat the one derived in the last subsection.
In particular, they only differ by an overall constant factor in the case without the NUT charge
and by further making identifications: $\alpha = \talpha\ta$, $m = \tm$, $a = \ta$, $r = \tr$,
$x = \tx$.

The final step that we need to recast the above solution into the familiar form of the general
P-D metric is to make the following linear combinations of the Killing coordinates:
\bea
t = \frac{\sqrt{\talpha\ta}(\tilde{t} -2\ta\tp)}{\sqrt{1
 +\talpha^2\big(\ta^2-\tn^2\big)^2}} \, , \qquad
\varphi = \frac{\ta\talpha^2\big(\ta^2-\tn^2\big)\tilde{t}
  +\big[1 -\talpha^2\big(\ta^4-\tn^4\big)\big]\tp}{\sqrt{\talpha\ta}
 \sqrt{1 +\talpha^2\big(\ta^2-\tn^2\big)^2}} \, , \nn
\eea
which yield the familiar expressions:
\bea
\tX = d\tilde{t} +\big[2\tn\tx -\ta\big(1-\tx^2\big)\big]d\tp \, , \quad
\tY = \ta\, d\tilde{t} -\big(\tr^2+\ta^2+\tn^2\big)d\tp \, . \nn
\eea
Note that in this new frame, the Rindler horizon is rotating relative to the conformal infinity.

To summarize, we have rederived the accelerating Kerr-NUT solution via the ISM procedure from
the Rindler background seed. Then using two different parametrizations, we have recast the
solution into two distinct forms in terms of the spherical-like coordinates, namely, the one
novelly given by Ref. \cite{PRD109-084038}, and another by the traditional one. Since the latter
is already known to be of type D, so is the one delivered by Astorino in \cite{PRD109-084038}.
However, one still needs to find what coordinate transformations may relate them. We leave
this subject to the next section.

By the way, here we would like to mention the related work \cite{PRD93-044021} that dealt with
the Euclidean P-D metric. In the Sec. VC of that paper, Chen obtained the Euclidean
accelerating Kerr-NUT metric as a 3-soliton solution via the ISM procedure. Then he performed a
M\"{o}bius transformation on the coordinates ($u, v$) and redefined the Killing coordinates, and
further made various parameter identifications so that the 3-soliton solution is brought to the
Euclidean P-D solution in the familiar form. This hints that the M\"{o}bius transformation may
be relevant to the present work.

\section{M\"{o}bius transformation
 amongst two metrics}\label{Mtct}

In order to find the relation between the above two metrics, a direct routine is to examine the
solutions given by Eqs. (\ref{Ast}) and (\ref{PD0}), respectively, since both of them adopt the
same Killing coordinates ($t, \varphi$). Therefore, the remaining most possible coordinate
transformation is the M\"{o}bius transformation amongst the radial and azimuthal coordinates,
respectively, together with further parameter identifications.

Let us consider the following M\"{o}bius transformation:
\be\label{Mt}
\tx = \frac{x -\nu}{1 -\nu\,x} \, , \qquad
\talpha\tr = \frac{\alpha\, r +\nu}{\ta(1+\nu\alpha\, r)-\tn(\alpha\, r +\nu)} \, ,
\ee
such that $\tx = \pm\, 1$ transform to $x = \pm\, 1$ under the first one, and the second one only
involves mapping between the acceleration horizons but not the event horizons among both metrics.
The inverse of the second transformation is
\be\label{ctr}
\alpha\, r = \frac{(\ta -\nu\tn)\talpha\tr -\nu}{1 +(\tn -\nu\ta)\talpha\tr} \, .
\ee
Here and hereafter we prefer to leave $\nu$ to be a freely variable parameter so that the relation
between the two mass parameters is left to be determined only in the final step.

We first apply the above transformation to $\tSigma = \tr^2 +(\tn +\ta\tx)^2$ and find its relation
to $\Sigma$. To do so, it is easy to find that if the identity $\tSigma(1 -\nu\,x)^2[\ta(1+\nu\alpha\,
r) -\tn(\alpha\, r +\nu)]^2 = C\,\Sigma$ holds true, then we must have the following conditions:
\be\begin{split}
& \frac{n}{a} = -\,\frac{\big(1+\nu^2\big)\tn -2\nu\ta}{\big(1+\nu^2\big)\ta -2\nu\tn} \, , \qquad
a^2C = \frac{\ta(\tn -\nu\ta)(\ta -\nu\tn)\big[\big(1+\nu^2\big)\ta -2\nu\tn\big]^2}{
 \big(1+\nu^2\big)\tn -2\nu\ta} \, , \\
& \talpha^2 = \frac{-\nu}{(\tn-\nu\ta)(\ta -\nu\tn)\big(\ta^2-\tn^2\big)} \, , \qquad
\alpha^2a^2 = \frac{-\nu(\ta-\nu\tn)\big[\big(1+\nu^2\big)\ta -2\nu\tn\big]^2}{
\big(1-\nu^2\big)^2(\tn-\nu\ta)\big(\ta^2-\tn^2\big)} \, . \label{anC}
\end{split}\ee
Next, we seek the links between the one-forms ($\tilde{X}, \tilde{Y}$) in (\ref{tXtY}) and their
counterparts ($X, Y$) without rescaling the Killing coordinates:
\bea
X &=& \big[1 -\big(a^2-n^2\big)\alpha^2x^2\big]\frac{dt}{\sqrt{\alpha}}
  +\big[2nx -a\big(1+x^2\big)\big]\sqrt{\alpha}d\varphi \, , \nn \\
Y &=& \big[a\big(1+\alpha^2r^2\big) +2n\alpha\,r\big]\frac{dt}{\sqrt{\alpha}}
  +\big(r^2+n^2-a^2\big)\sqrt{\alpha}d\varphi \, . \nn
\eea
It is relatively easy to examine $\sqrt{\talpha\ta}\sqrt{1 +\talpha^2\big(\ta^2-\tn^2\big)^2}$
($\tilde{X},\tilde{Y}$) rather than ($\tilde{X}, \tilde{Y}$). We further find
\be\begin{split}
\sqrt{\talpha\ta}\sqrt{1 +\talpha^2\big(\ta^2-\tn^2\big)^2}(1-\nu\,x)^2\tilde{X}
 &= \frac{\ta\sqrt{\alpha}\big(1-\nu^2\big)}{\ta -\nu\tn} X \, , \\
\sqrt{\talpha\ta}\sqrt{1 +\talpha^2\big(\ta^2-\tn^2\big)^2}[\ta(1+\nu\alpha\,r)
 -\tn(\alpha\, r +\nu)]^2\tilde{Y} &=
 -\frac{\ta^2\alpha^{3/2}\big(1-\nu^2\big)}{\talpha(\ta -\nu\tn)} Y \, , \label{XY}
\end{split}\ee
and must additionally impose
\be\label{aa}
a = -\,\frac{\talpha(\ta-\nu\tn)\big[\big(1+\nu^2\big)\ta
 -2\nu\tn\big]}{\alpha\big(1-\nu^2\big)} \, ,
\ee
then we can write
\be\label{nC}
n = \frac{\talpha(\ta-\nu\tn)\big[\big(1+\nu^2\big)\tn
 -2\nu\ta\big]}{\alpha\big(1-\nu^2\big)} \, , \quad
C = -\alpha^2\frac{\ta(\tn -\nu\ta)^2\big(\ta^2 -\tn^2\big)\big(1-\nu^2\big)^2}{\nu
\big[\big(1+\nu^2\big)\tn -2\nu\ta\big]} \, .
\ee
In practice, if we first find the links between ($\tilde{X}, \tilde{Y}$) and ($X, Y$) directly, then
all the above relations can be obtained, and it can be checked that the identity $\tSigma(1 -\nu\,x)^2
[\ta(1+\nu\alpha\, r) -\tn(\alpha\, r +\nu)]^2 = C\,\Sigma$ is also automatically fulfilled.

Note that the relation for $\alpha^2$ listed above then becomes the identity for $\talpha^2$, so
$\alpha$ can now temporarily be viewed as a free parameter too. Alternatively, we can also express
this in terms of untilded parameters as follows:
$$ \alpha^2 = \frac{\nu(a -\nu\,n)}{(n-\nu\,a)\big(a^2-n^2\big)} \, .$$

Our next step is to find the relation between the mass parameters. Using the following identities,
\bea
1 -\tx^2 = \frac{1-\nu^2}{(1-\nu\,x)^2}\big(1-x^2\big) \, , \quad
\frac{d\tx^2}{1 -\tx^2} = \frac{1-\nu^2}{(1-\nu\,x)^2}\,\frac{dx^2}{1 -x^2} \, , \nn
\eea
we can simply let
\bea
\frac{d\tx^2}{\tP(\tx)} = \big(1-\nu^2\big)\frac{dx^2}{P(x)} \, ,
\eea
so that after using the above relation for $\talpha^2$, we find that this is true only when we
further set
\be\label{mm}
\tm = \frac{\nu}{2\talpha(\ta-\nu\tn)} \, , \qquad
m = \frac{\nu\big(\nu^2\ta -3\ta +2\nu\tn\big)}{2\alpha(\ta-\nu\tn)} \, ,
\ee
which can also be attained by exploiting the radial part.

Finally, with the help of the following useful expressions,
\bea
1 +\talpha\tr(\tn +\ta\tx) = \frac{\ta\big(1-\nu^2\big)(1+\alpha\,rx)}{(1 -\nu\,x)
[\ta(1+\nu\alpha\,r) -\tn(\alpha\, r +\nu)]} \, , \quad
\tSigma = \frac{C\,\Sigma}{(1 -\nu\,x)^2[\ta(1+\nu\alpha\,r) -\tn(\alpha\, r +\nu)]^2} \, , \nn \\
\tD(\tr) = \frac{\big(1-\nu^2\big)\ta^2\alpha^2}{\talpha^2[\ta(1+\nu\alpha\,r)
 -\tn(\alpha\, r +\nu)]^4}\Delta(r) \, , \quad
\tP(\tx) = \frac{1-\nu^2}{(1-\nu\,x)^4}P(x) \, , \quad
\frac{d\tr^2}{\tD(\tr)} = \big(1-\nu^2\big)\frac{dr^2}{\Delta(r)} \, , \nn
\eea
and the relations (\ref{XY}) between ($\tX, \tY$) and ($X, Y$), we can show that the two
metrics only differ by an overall constant factor:
\bea
d\ts^2 &=& \frac{C\,\Sigma}{\big(1-\nu^2\big)\ta^2(1 +\alpha\,rx)^2}\Big[\frac{dr^2}{\Delta(r)}
 +\frac{dx^2}{P(x)}\Big] +\frac{\ta\alpha^3\big(1-\nu^2\big)}{C\talpha^3(\ta -\nu\tn)^2
 \big[1+\talpha^2\big(\ta^2-\tn^2\big)^2]}\,
 \frac{-\Delta(r)X^2 +P(x)Y^2}{(1 +\alpha\,rx)^2\Sigma} \nn \\
&=& \frac{C}{\ta^2\big(1-\nu^2\big)}\bigg\{
\frac{\Sigma}{(1 +\alpha\,rx)^2}\Big[\frac{dr^2}{\Delta(r)} +\frac{dx^2}{P(x)}\Big]
 +\frac{n}{(1-\nu^2)(\tn-\nu\ta)}\frac{-\Delta(r)X^2 +P(x)Y^2}{(1 +\alpha\,rx)^2\Sigma}\bigg\}
 \nn \\
&=& \frac{C\big(1+\alpha^2a^2\big)}{\ta^2\big(1-\nu^2\big)}\, ds^2 \, ,
\eea
provided that we set
$$1 +\alpha^2a^2 = \frac{n}{(1-\nu^2)(\tn-\nu\ta)}$$
which yields an equation to determine $\alpha$:
\be
\frac{\talpha}{\alpha}\big[\big(1+\nu^2\big)\tn-2\nu\ta\big]
 = \big(1-\nu^2\big)^2\frac{\tn-\nu\ta}{\ta-\nu\tn}
 -\frac{\nu}{\ta^2-\tn^2}[\big(1+\nu^2\big)\ta -2\nu\tn\big]^2 \, . \label{a1}
\ee

Thus we have generally proven that the two metrics given in the last two sections are isometric
to each other. It seems likely to let $C = \ta^2\big(1-\nu^2\big)/\big(1+\alpha^2a^2\big)$ so that
both metrics are completely identical. This condition implies
\be
\frac{\talpha^2}{\alpha^2} = \frac{\big(\nu^4+1\big)\ta^2 -\nu\big(3\nu^2+1\big)\ta\tn
 +\big(3\nu^2-1\big)\tn^2}{\ta(\ta -\nu\tn)\big(\ta^2-\tn^2\big)\big(1 -\nu^2\big)} \, .
\ee
When combined with Eq. (\ref{a1}), they yield a cubic equation for $\ta/\tn$:
\be
\nu(\nu^4-\nu^2+1)\ta^3 -\nu^2(3\nu^2-2)\ta^2\tn +\nu(3\nu^2-4)\ta\tn^2 +\tn^3 = 0 \, .
\ee

One can also conduct a similar task as done in the above by using the C-metric-like coordinates
($x, y$) with $y = -1/(\alpha\,r)$ and ($\bx, \by$) defined via $\by = -1/(\talpha\tr)$, $\bx =
\tn +\ta\tx$, rather than by employing the spherical-like coordinates. In this case, both metrics
become more symmetric, and now $\ta$ behaves like a scale parameter; for example, $\by = \tn
+\ta(y -\nu)/(1 -\nu\,y)$. Since the results remain the same as those given in the above, we
will not repeat this work although it might be relatively easy.

\subsection*{What limit should be taken in the
original P-D metric to get the type-D NUT C-metric?}

Previously, Griffiths and Podolsk\'{y} \cite{CQG22-3467} demonstrated that in the limit of zero
angular momentum parameter, the NUT C-metric is \emph{apparently} accelerating and can be transformed
into the Taub-NUT solution; as a result of this, the type-D NUT C-metric seems to be absent from
the most general family of the type-D P-D solution, leaving a puzzle for almost two decades. The
reason for this is probably that they did not consider the M\"{o}bius transformation. On the other
hand, it is a simple matter to just set $a = 0$ in the Astorino's expression \cite{PRD109-084038}
to arrive at the type-D C-metric, but this still does not explain how to get the type-D NUT C-metric
from the most general P-D solution.

To elucidate the lesson why the type-D NUT C-metric is missing from the P-D metric, let us
consider what limit we should take in the original P-D metric to get the expected type-D
NUT C-metric. Let us first assume that the transformation parameter $\nu$ and all the solution
parameters ($\tm, \tn, \ta, \talpha$) as well as the radial coordinate $\tr$ are all finite
and nonzero (although $\ta$ can be set to zero). To achieve the static NUT C-metric, we must
first let the rotation parameter $a$ tend to zero ($a\to 0$). Then from the first identity
in (\ref{anC}), the NUT charge parameter $n$ must also tend to a very tiny constant ($n\simeq
0$) so that the ratio $n/a$ keeps finite. From the fourth one in (\ref{anC}) or (\ref{aa}),
we know that the acceleration parameter $\alpha$ must approach the infinity ($\alpha\to\infty$)
so that $\alpha\,a$ is also finite. The second identity in (\ref{anC}) tells us that the overall
factor $C$ should tend to the infinity ($C\to\infty$) so that $C\alpha$ can be kept finite. From
the inverse M\"{o}bius transformation (\ref{ctr}), we can recognize that the radial coordinate
$r$ must approach zero so that $\alpha\,r$ remains finite. Finally, from the second equation
of Eq. (\ref{mm}), we learn that the mass parameter $m$ should tend to a very tiny constant
$(m\simeq 0)$. In the $a\to 0$ limit, both ($\alpha\,a$, $\alpha\,n$, $\alpha\,m$) and $\alpha\,r$
remains fixed.

To summarize, in order to get the type-D NUT C-metric, the corresponding limit is the following set:
\be
a\to 0\, , \quad n\simeq 0\, , \quad m\simeq 0\, , \quad r\to 0\, , \qquad
\alpha\to \infty \, ,
\ee
apart from the appropriate rescaling of the whole line element plus the linear combinations of
the Killing coordinates ($t\, ,\varphi$). However, the key measure is to consider the above
M\"{o}bius transformation (\ref{Mt}).

\section{Transforming the type-D NUT C-metric
into the familiar P-D form}\label{NUTCPD}

In this section, we will consider another obvious coordinate transformation that brings the
type-D NUT C-metric (\ref{NUTC}) into the familiar P-D form. To this end, let us introduce
the following coordinate transformations:
\be
t = 4n\mu(\tilde{t} +\tp) \, , \quad
\varphi = 4n\mu\alpha(\tilde{t} -\tp) \, , \quad
r = n\frac{q +1}{q -1} \, ,\quad  x = \frac{1-p}{n\alpha(1+p)} \, , \quad
m = \frac{n}{2\mu} \, ,
\ee
which bring the NUT C-metric (\ref{NUTC}) into the familiar P-D form:
\be
ds^2 = \frac{8n^2\mu}{(p -q)^2}\bigg\{\frac{Q(q)\big(d\tilde{t} +p^2d\tp\big)^2
 +P(p)\big(q^2d\tilde{t} -d\tp\big)^2}{1 +p^2q^2} +\big(1 +p^2q^2\big)\Big[
 \frac{dq^2}{-Q(q)} +\frac{dp^2}{P(p)}\Big]\bigg\} \, ,
\ee
where the structure function reads
\be
P(\xi) = \big[n^2\alpha^2(1+\xi)^2 -(1-\xi)^2\big]\big(1 +4\mu\xi -\xi^2\big) \, ,
\ee
in which $\alpha$ can be simply set to unity. The only nonzero Kretschmann invariant is
\bea
R^{abcd}R_{abcd} &=& \frac{3\alpha^4(p-q)^6}{\mu^2
 \tilde{\nu}^4(1+p^2q^2)^6}\big[2\mu\big(1-\tilde{\nu}^2\big)pq\big(p^2q^2-3\big)
 +\big(\tilde{\nu}^2+1\big)\big(3p^2q^2-1\big)\big] \nn \\
&&\qquad \times \big[2\mu\big(1-\tilde{\nu}^2\big)
 \big(3p^2q^2-1\big) -\big(\tilde{\nu}^2+1\big)pq\big(p^2q^2-3\big)\big] \, ,
\eea
where $\tilde{\nu} = n\alpha$. This Kretschmann invariant has a similar factorized structure
form as that of the Taub-NUT solution, but it does not coincide with the latter because the
former contains a conformal factor $(p-q)^6$. It demonstrates that the type-D NUT C-metric
cannot reduce to the Taub-NUT solution unless the acceleration parameter vanishes ($\alpha=0$).

From the above expression for the structure function, it is clear that this quartic function
is factorized in an entirely distinct manner from the traditional way. This perhaps is the
primary reason why the type-D NUT C-mtric is ``lost" from the most general P-D solution.

\section{Concluding remarks}

In this paper, we have shown that the type-D accelerating Kerr-NUT metric presented in Ref.
\cite{PRD109-084038} is not a new solution in the sense that it is diffeomorphic to the original
P-D solution, but it is still very convenient for further investigations in the nonrotating
case where the angular momentum parameter vanishes, namely, the type-D static accelerating
solution with a nonzero NUT charge since this solution is unavailable in the previous literature.
We have justified this point from two aspects. First, we have applied the ISM procedure to
generate the accelerating Kerr-NUT solution from the seed metric -- Rindler vacuum background,
which can be equivalently obtained as in Ref. \cite{PRD109-084038} via a limiting procedure
($a_2\to 0, l_2\to 0$ and then $w_4\to\infty$) on the double Kerr-NUT solution. Then we adopted two
different parametrizations to derive the novel metric given by Astorino in Ref. \cite{PRD109-084038}
and the traditional one early delivered by Griffiths and Podolsk\'{y}. This clearly demonstrates
that these two metrics are isometric, though displayed in two different forms. Second, we have
also provided the obvious coordinate transformations (i.e., M\"{o}bius transformation of the
radial and angular coordinates and linear combinations of the Killing coordinates) and parameter
identifications between these two line elements. Therefore, our work provides an explicit
verification of the insight of Ref. \cite{PRD109-084038} that the obtained spacetimes not only
are of type D, but also might be included into the P-D family. In particular, we also presented
another concrete M\"{o}bius transformation and the linear combinations of the Killing coordinates
that obviously cast the type-D NUT C-metric into the familiar form of the P-D solution. We anticipate
that this work can give a final answer to resolve the ``missing" puzzle of the type-D NUT C-metric.

In the Appendix \ref{Appe}, we have given two simple examples to illustrate our new normalization
scheme by deriving the five-dimensional Schwarzschild-Tangherlini solution from the flat vacuum via
the ISM construction. Our normalization strategy can be viewed as a methodological complement to
the Pomeransky \emph{``remove--re-add"} tricky.

\medskip
\textit{Note added}. Recently, an eprint \cite{2409.02308} appeared with a partial overlap of the
subject for which the M\"{o}bius transformation is also used to relate the Astorino's new form to
three different forms of the most general P-D solution. Our paper, however, focuses on the proof
that the Astorino's novel form can be reproduced from the already-known accelerating Kerr-NUT
solution and explains why the type-D NUT C-metric is ``missing" from the rotating and accelerating
P-D solution.

\section*{Acknowledgements}

This work is supported by the National Natural Science Foundation of China (NSFC) under Grants
No. 12375053, No. 12205243, and No. 11675130, by the Sichuan Science and Technology Program under
Grant No. 2023NSFSC1347, and by the Doctoral Research Initiation Project of China West Normal
University under Grant No. 21E028.

\section*{Appendix: ISM construction of five-dimensional static
solutions with our new normalization scheme}\label{Appe}

In this Appendix, we will show that by using our normalization scheme, the five-dimensional
Schwarzschild-Tangherlini solution can be generated from the flat vacuum background via the
ISM construction, contrary to the statement made in Ref. \cite{PRD73-044004}. Our derivation
strictly follows the spirit of the ISM construction, and is different from what had been done
in Ref. \cite{PTP114-793} where the ansatz of the Killing metric is modified.

\setcounter{equation}{0}
\renewcommand{\theequation}{A.\arabic{equation}}
\subsection{Generating five-dimensional Schwarzschild-Tangherlini solution}

We first take the five-dimensional flat Minkowski background
\bea
ds^2 = -dt^2 +\rho^2d\phi^2 +d\psi^2 +d\rho^2 +dz^2 \, , \nn
\eea
as our seed metric:
\be
\bg_0 = \diag(-1\, , \rho^2 \, , 1) \, ,
\ee
so the generating matrix is easily written as
\bea
\bPsi_0 = \diag(-1\, , \rho^2 -2\lambda\,z -\lambda^2 \, , 1) \, . \nn
\eea

We now perform a 2-soliton transformation: add a soliton $\mu_1$ at $z = z_1$ with the vector
$m_0^{(1)} = (-1\, , 0\, , 0)$ and a soliton $\mu_2$ at $z = z_2$ with the vector $m_0^{(2)} =
(0\, , -2z_2\, , 0)$, respectively. Then two BZ vectors constitute a $2\times\, 3$ matrix
and $\bGa$ is a $2\times\, 2$ diagonal matrix:
\bea
\bm = \bigg(\begin{array}{ccc}
 1 & 0 & 0 \\
 0 & \frac{1}{\mu_2} & 0
\end{array} \bigg) \, , \qquad
\bl = \bigg(\begin{array}{ccc}
 -1 & 0 & 0 \\
 0 & \frac{\rho^2}{\mu_2} & 0
\end{array} \bigg) \, , \qquad
\bGa = \bigg(\begin{array}{cc}
\frac{-1}{\rho^2 +\mu_1^2} & 0 \\
0 & \frac{\rho^2}{\mu_2^2\big(\rho^2 +\mu_1^2\big)}
\end{array} \bigg) \, . \nn
\eea

The generated five-dimensional static 2-soliton metric is very simple:
\be
\widetilde{\bg} = \diag\Big(\frac{\rho^2}{\mu_1^2} \, , \frac{-\rho^4}{\mu_2^2} \, , 1\Big) \, ,
\ee
with the determinant $\det\widetilde{\bg} = -\rho^6/\big(\mu_1^2\mu_2^2\big) \not= -\rho^2$.

It is suggested to take the normalization diagonal matrix as
\be
\bW = \diag\Big(\frac{\mu_1^{3/2}}{\rho\sqrt{\mu_2}} \, , \frac{\mu_2}{\rho\sqrt{\mu_1}} \, ,
\sqrt{-\mu_2}\Big) \, ,
\ee
then the final physics metric is exactly the expected one:
\be
\bg = -\bW\widetilde{\bg}\bW = -\bW^2\widetilde{\bg} =
\diag\Big(\frac{-\mu_1}{\mu_2} \, , \frac{\rho^2}{\mu_1} \, , \mu_2\Big) \, ,
\ee
with the conformal factor being recorded as
\be
f = \frac{\mu_2\big(\rho^2+\mu_1\mu_2\big)}{\big(\rho^2+\mu_1^2\big)\big(\rho^2+\mu_2^2\big)} \, .
\ee

This trivial example illuminates that the five-dimensional Schwarzschild-Tangherlini solution is
feasibly generated via the ISM procedure with our new normalization scheme from the flat Minkowski
vacuum seed.

\setcounter{equation}{0}
\renewcommand{\theequation}{B.\arabic{equation}}
\subsection{Generating five-dimensional static Emparan-Reall black ring}

Next, we want to rederive the static Emparan-Reall black ring \cite{PRL88-101101} solution
from the five-dimensional flat Minkowski background written in another form:
\be
ds^2 = -dt^2 +\frac{\rho^2}{\mu_3}d\phi^2 +\mu_3d\psi^2
 +\frac{\mu_3}{\rho^2+\mu_3^2}\big(d\rho^2 +dz^2\big) \, .
\ee
Clearly, its $t = {\rm\, const}$ slice is the Euclidean Rindler metric.

From the seed metric
\be
\bg_0 = \diag\Big(-1 \, ,\frac{\rho^2}{\mu_3} \, , \mu_3\Big) \, ,
\ee
we rewrite $\rho^2/\mu_3 = -\nu_3$ and make the replacements, $\mu_3\to \mu_3 -\lambda$,
$\nu_3\to \nu_3 -\lambda$, then easily obtain the generating matrix:
\bea
\bPsi_0 = \diag\Big(-1 \, ,\frac{\rho^2}{\mu_3} +\lambda \, , \mu_3 -\lambda\Big) \, . \nn
\eea

We now perform a 2-soliton transformation: add a soliton $\mu_1$ at $z = z_1$ with the vector
$m_0^{(1)} = (1\, , 0\, , 0)$ and a soliton $\mu_2$ at $z = z_2$ with the vector $m_0^{(2)} =
(0\, , 1 \, , 0)$, respectively. The order for the 3-soliton is aligned as $z_1 < z_2 < z_3$.
Then two BZ vectors constitute a $2\times\, 3$ matrix and $\bGa$ is also a diagonal $2\times\,
2$ matrix:
\be
\bm = \bigg(\begin{array}{ccc}
 -1 & 0 & 0 \\
 0 & \frac{\mu_3}{\rho^2 +\mu_2\mu_3} & 0
\end{array} \bigg) \, , \quad
\bl = \bigg(\begin{array}{ccc}
 1 & 0 & 0 \\
 0 & \frac{\rho^2}{\rho^2 +\mu_2\mu_3} & 0
\end{array} \bigg) \, , \quad
\bGa = \bigg(\begin{array}{cc}
\frac{-1}{\rho^2 +\mu_1^2} & 0 \\
0 & \frac{\rho^2\mu_3}{\big(\rho^2 +\mu_2^2\big)\big(\rho^2+\mu_2\mu_3\big)^2}
\end{array} \bigg) \, . \nn
\ee

The generated five-dimensional 2-soliton metric is still diagonal:
\be
\widetilde{\bg} = \diag\Big(\frac{\rho^2}{\mu_1^2} \, ,
\frac{-\rho^4}{\mu_2^2\mu_3} \, , \mu_3\Big) \, ,
\ee
with the determinant $\det\widetilde{\bg} = -\rho^6/\big(\mu_1^2\mu_2^2\big) \not= -\rho^2$.

Now we suggest to take the normalization matrix as
\be
\bW = \diag\Big(\frac{\mu_1^{3/2}}{\rho\sqrt{\mu_2}} \, , \frac{\mu_2^{3/2}}{\rho\sqrt{\mu_1}} \, ,
\sqrt{-1}\Big) \, ,
\ee
then the final physics metric is exactly that of the static Emparan-Reall black ring:
\be
\bg = -\bW\widetilde{\bg}\bW = -\bW^2\widetilde{\bg} =
\diag\Big(\frac{-\mu_1}{\mu_2} \, , \frac{\rho^2\mu_2}{\mu_1\mu_3} \, , \mu_3\Big) \, ,
\ee
whose conformal factor can be easily computed as
\be
f = \frac{k\, \mu_3\big(\rho^2+\mu_1\mu_2\big)^2\big(\rho^2+\mu_2\mu_3\big)}{(1-\nu)
\big(\rho^2+\mu_1\mu_3\big)\big(\rho^2+\mu_1^2\big)\big(\rho^2+\mu_2^2\big)\big(\rho^2+\mu_3^2\big)}
\, .
\ee

In the coalescing limit $\mu_3\mapsto \mu_2$ with $k = 1 -\nu$, the solution reduces to the
five-dimensional Schwarzschild-Tangherlini solution. On the other hand, in the merging limit
$\mu_2\mapsto \mu_1$ with $k = 1 -\nu$, the solution coalesces to the above five-dimensional
Minkowski background metric.

\def\CQG{Classical Quantum Gravity\,}
\def\JHEP{J. High Energy Phys.\,}
\def\PRD{Phys. Rev. D\,}
\def\PRL{Phys. Rev. Lett.\,}
\def\PLA{Phys. Lett. A\,}
\def\PLB{Phys. Lett. B\,}
\def\AP{Ann. Phys. (N.Y.)\,}
\def\JMP{J. Math. Phys. (N.Y.)\,}
\def\PTP{Prog. Theor. Phys.\,}

\end{document}